\documentclass{aa}

\usepackage{graphicx}
\usepackage{natbib}
\bibpunct{(}{)}{;}{a}{}{,}
\begin{document}
\newcommand{\block}{\vrule width 0.5 true cm height 6pt depth 0pt \ }
\newcommand{\vc}[1]{\mbox{\protect\boldmath$#1$}}

  \title{Ram pressure stripping of tilted galaxies}
%  \title{Stripping effects of ram pressure on tilted galaxies}

    \author{
        P. J\'achym     \inst{1}
    \and
        J. K\"oppen     \inst{1,2,3,4}
    \and
        J. Palou\v s    \inst{1}
    \and
        F. Combes       \inst{5}
           }
    \offprints{P. J\'achym }

    \institute{Astronomical Institute,
           Academy of Sciences of the Czech Republic,
           Bo\v cn\' \i \ II 1401, 141 31 Prague 4, Czech Republic\\
           \email{jachym@ig.cas.cz, palous@ig.cas.cz}
          \and
            Observatoire Astronomique de Strasbourg,
            11 Rue de l'Universit\'e,
            67000 Strasbourg, France\\
            \email{koppen@astro.u-strasbg.fr}
         \and
            International Space University,
            Parc d'Innovation,
            1 Rue Jean-Dominique Cassini,
            F--67400 Illkirch-Graffenstaden, France
         \and
            Institut f\"ur Theoretische Physik und Astrophysik,
            Universit\"at Kiel,
            D--24098 Kiel, Germany
          \and
            Observatoire de Paris, LERMA,
            61 Av. de l'Observatoire,
            75014 Paris, France\\
            \email{francoise.combes@obspm.fr}
        }

   \date{Received 4 December, 2008 / accepted 30 March, 2009}

\titlerunning{Ram pressure stripping of tilted galaxies}
\authorrunning{J\'achym et al.}

\abstract
{Ram pressure stripping of galaxies in clusters can yield gas deficient 
disks. Previous numerical simulations based on various approaches 
suggested that, except for near edge-on disk orientations, the amount of 
stripping depends very little on the inclination angle.}
{Following our previous numerical and analytical study of face-on
stripping, we extend the set of parameters with the disk tilt angle and 
explore in detail the effects of the ram pressure on the interstellar 
content (ISM) of tilted galaxies that orbit in various environments of 
clusters, with compact or extended distributions of the intra-cluster 
medium (ICM). We further study how results of numerical simulations 
could be estimated analytically. To isolate the effect of inclination, 
galaxies on strictly radial orbits are considered.}
{A grid of numerical simulations with varying parameters is produced 
using the tree/SPH code GADGET with a modified method for calculating 
the ISM-ICM interaction. These SPH calculations extend the set of 
existing results obtained from different codes using various numerical 
techniques.}
{The simulations confirm the general trend of less stripping at 
orientations close to edge-on. The dependence on the disk tilt angle 
is more pronounced for compact ICM distributions, however it almost 
vanishes for strong ram pressure pulses. Although various 
hydrodynamical effects are present in the ISM-ICM interaction, the main 
quantitative stripping results appear to be roughly consistent with a 
simple scenario of momentum transfer from the encountered ICM. This 
behavior can also be found in previous simulations. To reproduce the 
numerical results we propose a fitting formula depending on the disk 
tilt angle and on the column density of the encountered ICM. Such a 
dependence is superior to that on the peak ram pressure used in 
previous simple estimates. }
{}

\keywords{Galaxies: clusters: general -- Galaxies: evolution -- Galaxies: 
intergalactic medium -- Galaxies: ISM -- Galaxies: structure}

\maketitle
%________________________________________________________________
\section{Introduction}
Ram pressure of the inter-galactic medium in the dense environments of 
galaxy clusters and groups can significantly affect the interstellar 
content of galaxies. Together with tidal stripping and gas outflows due 
to galaxy mergers, ram pressure stripping is an efficient mechanism  
removing their gas component. This quenches the star formation. Local 
enhancement of the gas density by the ram pressure, however, can 
increase the star formation rate in the central disk regions and in the 
tail of stripped material \citep{kronberger08}. Current observations 
showing gaseous tails and extra-planar gas 
\citep[e.g.][]{crowl05,chung07}, and a number of numerical simulations 
\citep[e.g.][]{vollmer01a, roedigerbruggen07} confirm the importance of 
gas stripping in the evolution of galaxies. Although some gas tails 
could be due to tidal interactions between galaxies in the cluster 
\citep{chung07}, these are rapidly destroyed by the ram pressure 
\citep{kapferer08}. Ram pressure then creates its own long gas tails 
\citep{roedigerbruggen08}.

When the galaxy plunges into the intracluster medium (ICM), a complex 
hydrodynamical interaction of the interstellar matter (ISM) and ICM 
takes place. Several phases of ISM-ICM interaction were recognized by 
\citet{roedigerbruggen06}: the initial ram pushing of the ISM out of 
the disk, followed by stripping of elements accelerated to the escape 
velocity, in a background supported by continuous stripping. A simple 
estimate of the stripping amount suggests that if the ram pressure of 
the wind exceeds the local gravitational restoring force per unit area, 
the gas is released from the potential of the galaxy 
\citep[][GG72]{gunngott72}. This approximation works well for face-on 
cases with constant long-lasting flows of the ICM. However, in real 
situations, galaxies experience a time-varying ram pressure due to 
their evolving orbital velocity and radially changing ICM density. An 
approximative criterion where the stripping is proportional to the 
momentum of the ICM encountered along the orbit was introduced by 
\citet{jachym07}.

A long-term ram pressure corresponding to the overall cluster ICM 
density profile can be accompanied by short pressure spikes from local 
clumps in the ICM distribution or wakes caused by previous stripping 
events. \citet{tonnesenbryan08} studied the role of ICM 
overdensities and their infalling motions, and suggested that even 
small substructures in the ICM may evoke a stronger stripping than that 
corresponding to the given position in the cluster. In short ICM-ISM 
encounters, not all the ISM is accelerated to the escape speed and its 
re-accretion can occur. Very recently, \citet{tonnesenbryan09} took 
into account the multi-phase nature of the ISM, and found different 
results, according to the clumpiness of the gas. \citet{vollmer09} 
argue that ram pressure is more efficient on ionized gas, and they see 
evidence of differential ram pressure efficiency according to the 
nature of the multiphase gas.

The question of how the orientation of the galaxy relative to the 
orbital direction influences the final stripping was touched on by many 
authors \citep{quilis00, vollmer01a, schulzstruck01, hidakasofue02, 
marcolini03, roedigerbruggen06}. For tilted galaxies the effect of the 
hydrodynamical shielding is more pronounced in comparison to the 
face-on case and yields weaker stripping. Moreover, due to galactic 
rotation, near edge-on ram pressures evoke asymmetric shapes of the ISM 
disks \citep[e.g.][]{roedigerbruggen06}.
From previous simulations, only close-to edge-on orientations were 
identified as less efficient in the stripping than face-on cases 
\citep{quilis00,roedigerbruggen06}. The tilted stripping was found to 
be delayed compared to face-on \citep{vollmer01a}, and takes a 
longer time \citep{faroukishapiro80}. Other features, like tails of 
stripped material forming on the disk side where the gas rotates 
against the wind \citep{phookunmundy95}, or annealing of inclined 
galaxies by a stronger loss of angular momentum \citep{schulzstruck01}, 
were noticed. Mimicking the hydrodynamical shielding of tilted galaxies 
with a simple geometrical prescription, \citet{vollmer01a} performed a 
set of sticky-particle simulations with time-varying ram pressure 
peaks. \citet{roedigerbruggen06,roedigerbruggen07}, using 
hydrodynamical adaptive-mesh-refinement simulations, introduced models 
of galaxies experiencing a constant ICM flow in a periodic boundary 
box, or orbiting on realistic orbits in clusters. 

What is the role of the galactic disk tilt angle in various 
distributions of the ICM, and could a simple dependence between the 
stripping amount and the tilt angle be found from simulations? To 
answer these questions we perform in this paper an extensive grid of 
Smoothed particle hydrodynamics (SPH) simulations with time-varying 
large or small ram pressure peaks of 
generally tilted ram pressure stripping events. This completes other 
existing simulations based on different numerical approaches (AMR, 
sticky-particles etc.) and follows our simulations on the face-on study 
made by \citet{jachym07}. 
Rather than describing in detail the process of stripping itself (that 
shall be studied in a forthcoming paper), we focus on the quantitative 
outcomes of the tilted ISM-ICM interactions. To isolate the effect of 
the disk inclination, we deal with galaxies on strictly radial orbits 
freely falling from the cluster outskirts. This condition however does 
not imply a strong limitation since ram pressure profiles experienced 
by galaxies along slightly elliptical orbits may be modeled with 
strictly-radial ones through an ICM distribution with a lower value of 
the central density and a lower concentration towards the center.
With the results of the simulation grid in hand, we use a simplified 
test particle model to separate the various effects taking place in the 
ISM-ICM interactions, and to search for a simple dependence 
that would fit the numerical results.

Our model, initial conditions and simulation method are introduced in 
Section~\ref{sec:model}. The results of simulations are presented and 
analyzed in Sections~\ref{sec:proc} and \ref{sec:res}. Their comparison 
with a simple test model is discussed in Section~\ref{sec:kin}. How can 
the results of the tilted stripping be estimated? This question is 
answered in Section~\ref{sec:analyt}, which deals with a momentum 
delivery criterion. In the discussion of Section~\ref{sec:disc}, we 
compare our results with outcomes of previous numerical simulations.

\section{Modeling}\label{sec:model}
\subsection{Galaxy model}
Our model galaxy is a late-type massive galaxy (LM-type in
\citealp{jachym07}): a three-component system with Plummer bulge
and dark halo, and with Toomre-Kuzmin-sech$^2(z/z_0)$ disk. The 
values of the model parameters are $M_d=8.6\ 10^{10}$ M$_\odot$, 
$a_d=4$ kpc, $z_0=0.25$ kpc, $M_b=1.3\ 10^{10}$ M$_\odot$, $a_b=0.4$ 
kpc, and $M_h=4.2\ 10^{11}$ M$_\odot$, $a_h=20$ kpc. All three 
components are represented with particles (12~000 for the stellar disk, 
12~000 for the halo, and 6000 for the bulge). Their 
distribution is truncated at 16 kpc (disk), 4 kpc (bulge), and 40 kpc 
(halo). The corresponding rotation curve is flat with a velocity of 
about 250~km~s$^{-1}$. The ISM in the disk follows the same density 
profile as the stellar component with the same radial scale ($a_d=4$ 
kpc). Its mass forms 10\% of the total disk mass. The ISM disk 
consists of 12~000 SPH particles which provides sufficient resolution 
for the large-scale effects of interest in this paper.

\subsection{Cluster model}
The galaxy cluster is modeled with dark matter (DM) and ICM gas. Their
volume densities follow a $\beta$-profile \citep{cavaliere76,
schindler99}: $\rho = \rho_0 (1 + R^2/ R_{\rm c}^2 )^{-3\beta /2}$.
For the {\it standard cluster model} we use $\beta_{\rm ICM}=1/2$, 
$\rho_{\rm 0,ICM}=6.5\ 10^{-3}$ cm$^{-3} = 0.64\ 10^{-26}$ g~cm$^{-3}$ 
and $R_{\rm c,ICM} = 13.4$ kpc. The fixed DM, which provides the 
gravitational potential, has $\rho_{\rm 0,DM}=3.8\ 10^{-4}$ 
M$_\odot\,$pc$^{-3}$,$R_{\rm c,DM} = 320$ kpc, and $\beta_{\rm DM}=1$. 
The ICM is introduced with 120~000 SPH particles distributed within 
a 140~kpc truncation radius about the cluster center. Initially, the 
ICM particles are assigned velocity dispersions calculated from the 
hydrostatic equilibrium equation of the cluster. The system then 
slightly relaxes to a stable state. Particles that reach the truncation 
radius of 140 kpc are relocated to the centrally symmetric position. 
See \citet{jachym07} for more details. 

\begin{table}[t]
\centering
\caption{\small
Set of basic simulation runs. We refer to $R1\rho1$ as to the
{\it standard run}. Values of $\Sigma_{\rm ICM}$ correspond to $R_{\rm 
ICM}^{\rm trunc}=140\ $kpc.}
\begin{tabular}{lcccc}
\hline
\hline \rule{-3.pt}{2.6ex}
run & $R_{\rm c,ICM}$, $\rho_{\rm 0,ICM}$ & $p_{\rm ram}^{\rm
max}$ & $\Sigma_{\rm ICM}$\\
& (kpc,10$^{-3}$cm$^{-3}$) & cm$^{-3}$(km/s)$^2$ & ($M_\odot$pc$^{-2}$)\\
\hline \rule{-3.pt}{2.6ex}
$R4\rho8$&53.6 , 52   & 100~913& 114\\
$R4\rho4$&53.6 , 26   & 46~752 &  57\\
$R4\rho1$&53.6 , 6.5  & 10~993 &  14\\
$R4\rho0$&53.6 , 1.6  & 2705   & 3.6\\
\hline \rule{-3.pt}{2.6ex}
$R1\rho8$&13.4 , 52   & 89~214 &  41\\
$R1\rho4$&13.4 , 26   & 43~827 &  20\\
$R1\rho1$&13.4 , 6.5  & 10~810 &   5\\
$R1\rho0$&13.4 , 1.6  & 2693   & 1.3\\
\hline \rule{-3.pt}{2.6ex}
$R0\rho8$&3.4  , 52   & 86~549 &  12\\
$R0\rho4$&3.4  , 26   & 43~143 & 5.9\\
$R0\rho1$&3.4  , 6.5  & 10~769 & 1.5\\
$R0\rho0$&3.4  , 1.6  & 2689   & 0.4\\
\end{tabular}
\label{tab_runs}
\end{table}

\subsection{Initial conditions}
The model galaxy initially starts with zero velocity from the cluster
periphery ($R=1$ Mpc) and freely falls along a strictly radial orbit
towards the cluster center. In the standard model the galaxy reaches
the center at time $T=1.64$ Gyr with a velocity of about 1300 
km~s$^{-1}$. The simulation is followed until $T=2$ Gyr 
when the galaxy has already left the central part of the cluster 
filled with ICM particles. Thus, each simulation run corresponds to one 
orbit of the galaxy through the cluster center.

To treat various cluster environments, we vary the values of the 
$R_{\rm c,ICM}$ and $\rho_{\rm 0,ICM}$ parameters, multiplying each 
standard value by factors of 8, 4, 1, and 0.25. Table~\ref{tab_runs} 
defines the basic set of simulation runs which are labeled with the 
corresponding factors (0 stands for the factor of 0.25). 
Table~\ref{tab_runs} gives for each run the combination of $R_{\rm 
c,ICM}$ and $\rho_{\rm 0,ICM}$ parameters, the peak value of the ram 
pressure along the orbit, and the column density of the ICM encountered
on the orbit. Note that narrow ICM distributions or those with low 
values of density may represent ICM overdensities or debris structures 
left over in the cluster from recent stripping events. 

\citet{abadi99} warned against too massive ICM particles that can punch 
holes in the gas disk and cause a large artificial drag on the ISM 
particles. In our standard model ($R1\rho1$) one ICM particle has a 
mass of about 5~10$^5$~M$_\odot$, which is almost equal to the mass of 
an individual ISM particle. Concerning the ICM component setup, its 
truncation radius of 140~kpc was chosen so that ram pressure in the 
standard cluster exceeds at that distance the gravitational restoring 
force in the disk outskirts of an approaching galaxy. 

We introduce the tilt angle of the disk with respect to the orbital 
direction of the galaxy as $i=0^\circ$ for edge-on and $i=90^\circ$ for 
face-on orientation, throughout the paper.

\subsection{Simulation method}\label{sec:method}
The simulations were carried out with the tree/SPH code GADGET
\citep{springel01} adapted by \citet{jachym07} for calculations of 
ISM-ICM interactions. Smoothed particle hydrodynamics 
\citep[SPH,][]{lucy77,gingoldmonaghan77} in its standard formulation
has significant problems with contact discontinuities where the density
jump is very large. Across such contact discontinuities, fluid
instabilities can be suppressed \citep[e.g.][]{agertz07}, or other 
artifacts can occur when the sampling is poor 
\citep[e.g.][]{okamoto03}. The problem of ICM-ISM interaction is 
basically a special case of this. \citet{jachym07} modified GADGET to 
estimate smoothing lengths of either ICM or ISM particles separately 
from neighbors of a corresponding phase. 
This means that the search of the neighboring particles continues 
until a desired number of proper neighbors are found, while ignoring 
those of the second phase. It ensures that even a moderate number of 
ICM particles will fully cover during the interaction the ISM disk, as 
the ICM particles do not shrink when approaching the dense ISM in the 
disk. This modified code was used by \citet{jachym07} for 
simulations of face-on ram pressure stripping events.

Although large ICM particles result in a low spatial resolution 
and ISM particles that lack pressure gradients from the wind, 
our suggestion provides a simple way to avoid the above limitation 
of the standard SPH method, and it may be used for the quantitative 
purposes of this paper. An ideal solution could involve a new 
self-consistent way to estimate densities that allows a better 
representation of contact discontinuities -- e.g. based on a 
tessellation scheme \citep[recently published by][]{springel09}, 
implementation of the particle splitting mechanism 
\citep{kitsionaswhitworth07} into the standard SPH, or the new SPH scheme 
of \citet{kawata09}.

\begin{figure}[t]
\centering
\includegraphics[height=0.48\textwidth,angle=270]{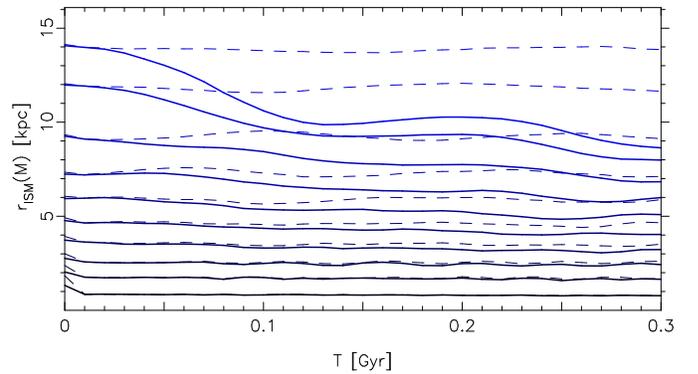}
\caption{\small 
Zero-order test vs. isolated evolution: Disk radii encompassing 95~\%, 
90~\%, 80~\%, ..., 10~\% of the total ISM mass ({\it from top down}) as 
functions of time in the case of the galaxy evolving at rest in the 
center of the standard cluster ({\it solid curves}), and in an isolated 
galaxy ({\it dashed}).
}\label{zero_test}
\end{figure}

\begin{figure*}[t]
\centering
\includegraphics[height=1.\textwidth,angle=270]{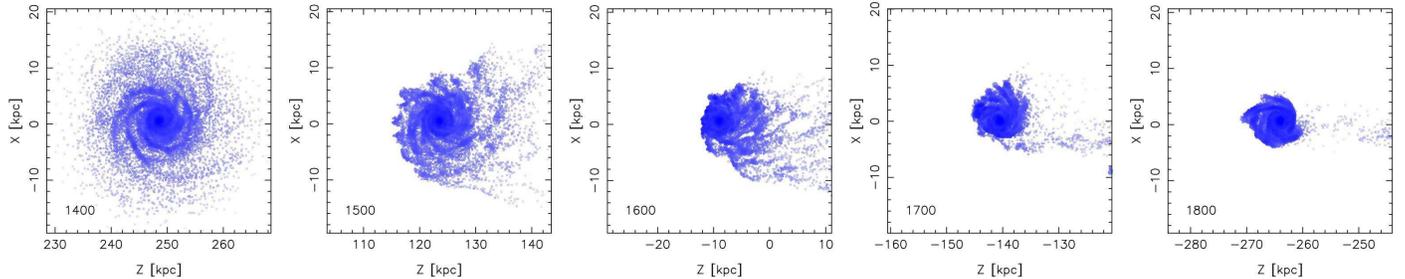}
\caption{\small Pole-on views of the ISM disk crossing edge-on through 
the $R4\rho1$ cluster. The disk rotates clockwise; the ram pressure 
operates from the left. Color saturation corresponds to increasing 
ISM density.}
\label{edge_snap}
\end{figure*}

\subsection{Zero-order simulation run}\label{zero}
In order to test the influence of the simulation technique used, and 
the effect of resolution on the results of our stripping simulations, 
we present a zero-order test: The galaxy is placed at rest at the 
center of the cluster and left to evolve surrounded by the ICM for about 
300 Myr (the time comparable to the duration of the standard ISM-ICM 
interaction discussed later). The evolution of the galaxy's gaseous 
disk is displayed in Fig.~\ref{zero_test} showing radii that 
encompass fixed fractions of the total ISM mass (solid curves) 
evolving with time. The ISM disk (or at least its outer layers) is 
compressed by an abrupt introduction of the static external pressure of 
the ICM on a time scale of about 100~Myr. After this period the 
outer layers start to slightly oscillate while the disk slowly 
continues to settle towards an equilibrium. All the ISM particles 
during the simulation stay bound to the galaxy. The stellar disk 
component is unchanged during the simulation. For comparison, the 
evolution of the ISM disk of an isolated galaxy, not exposed to any 
external pressure, is plotted in Fig.~\ref{zero_test} with dashed 
curves. 

The ISM component is treated isothermally in our simulations. This 
simple approach is used to roughly mimic an efficient radiative cooling 
of the ISM gas. The disk confinement in Fig.~\ref{zero_test} may thus 
be understood in terms of an abrupt increase of the external pressure 
after the galaxy has been placed into the ICM -- due to its 
isothermality the ISM cannot heat up (to compensate the increased 
external pressure) and is compressed instead. In Fig.~\ref{zero_test}, 
an extreme effect is shown due to a maximum pressure corresponding to 
the cluster center. However, in the ram pressure stripping simulations 
that we deal with further in this paper, the galaxy arrives at the 
cluster center only gradually from its outskirts. This aspect will 
be treated in the following section. 

Concerning the influence of our numerical technique on the simulation 
outputs, no Rayleigh-Taylor instabilities or spurious effects due to 
the lack of resolution appeared in the zero-order test simulation.

\section{Process of stripping}\label{sec:proc}
\begin{figure}[t]
\centering
\includegraphics[height=0.48\textwidth,angle=270]{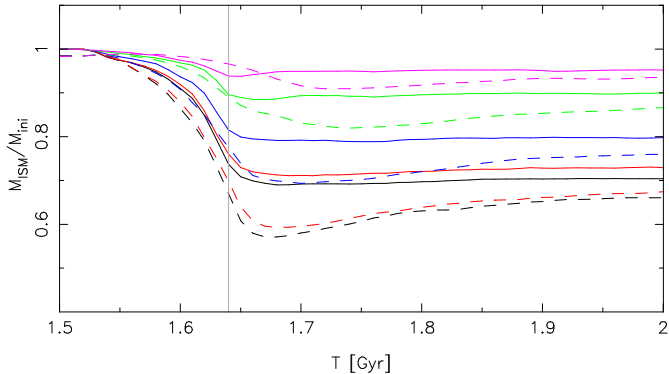}
\caption{\small 
Effect of the inclination: Evolution of the ISM mass bound to the 
galaxy ({\it solid curves}) or enclosed within an evaluation cylinder 
around the disk plane ({\it dashed}) in simulations with disk tilt 
decreasing from face-on to edge-on ({\it from bottom up}) of the 
standard $R1\rho1$ run.
}\label{M_tilt}
\end{figure}

\begin{figure}[t]
\centering
\includegraphics[height=0.48\textwidth,angle=270]{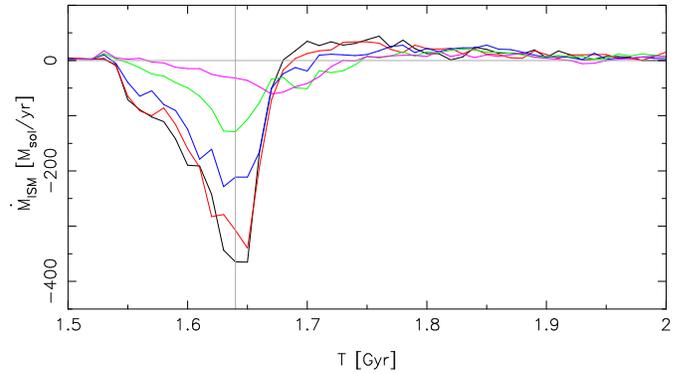}
\caption{\small 
Mass flow of the ISM through the boundary of the evaluation cylinder 
for five simulations of Fig.~\ref{M_tilt}. 
}\label{Mdot}
\end{figure}

For non-face-on orientations of the galaxy with respect to the ICM 
wind, the galactic rotation starts to play a role. Its effect is most 
pronounced for near edge-on cases when it evokes an asymmetry in 
the gaseous disk. This can be seen in Fig.~\ref{edge_snap} where 
snapshots of the galaxy passing edge-on through the center of a wide 
$R4\rho1$ cluster are depicted. (For illustrative purposes the 
$R4\rho1$ run is studied here rather than the standard one where 
edge-on stripping is less prominent.) Due to rotation, the disk side 
moving parallel to the wind (upper side in Fig.~\ref{edge_snap}) 
experiences ram pressure that is lower by a factor of $(v-v_{\rm 
rot})^2 / (v+v_{\rm rot})^2$ with respect to the opposite disk side; 
$v$ is the orbital speed of the galaxy and $v_{\rm rot}$ is the 
rotational velocity of the disk. However, as e.g. the second panel of 
the Figure shows, the parallel-rotating side is more severely stripped 
than the opposite one and it is thus not the highest ram pressure that 
yields the strongest stripping. This paradox was noted by 
\citet{roedigerbruggen06} as well. The explanation is in terms of time 
delay: on the parallel-rotating side, the gas is accelerated towards 
more rotational velocity and exits the galaxy almost at once. At the 
opposite side, the gas is being braked from $v-v_{\rm rot}$ towards $v$ 
and with lower or no centrifugal force it has time to fall towards the 
center of galaxy. Then, it still has to rotate half a turn around the 
galaxy, and it has time to be re-accreted (or at least it stays around 
the galaxy for a longer time).
Snapshots in Fig.~\ref{edge_snap} further show how, due to rotation, 
the tail of the stripped material winds up around the disk.

To measure the actual level of stripping we set up a cylindrical zone
about the galactic disk plane ($r=16$ kpc and $|z|\leq3$ kpc), and 
follow the enclosed ISM mass and the flow of the material through the 
zone boundary. This is shown in Figs.~\ref{M_tilt} and \ref{Mdot} for 
five simulations of the standard run with different tilt angles, 
together with the ISM mass fraction bound to the galaxy (solid curves 
in Fig.~\ref{M_tilt}). It confirms the expected trend: with the tilt 
angle growing towards face-on, more material is stripped. The maximum 
departure of the curves of the bound and in-the-zone-enclosed ISM mass 
in Fig.~\ref{M_tilt} indicates the amount of material that is 
re-accreted after the ISM-ICM interaction. The bound mass in 
Fig.~\ref{M_tilt} is given by the sum of the ISM particles for 
which $E_{\rm kin}+E_{\rm pot}<0$. Note that the system of the galaxy 
is not closed -- it experiences a changing effect of the external 
potential of the cluster and an external pressure. A slight growth of 
the solid curves in Fig.~\ref{M_tilt} may be found at later simulation 
times.

The "stripping rate" $\dot{M}_{\rm ISM}$, i.e. the flow of the ISM 
through the boundary of the evaluation zone, exceeds almost 400 
M$_\odot$yr$^{-1}$ for a face-on galaxy, the peak value decreases towards 
the edge-on orientation to about 50 M$_\odot$yr$^{-1}$ (see 
Fig.~\ref{Mdot}). This Figure shows not only the removal rate 
($\dot{M}_{\rm ISM}<0$) but also a reversed re-accretion rate of the 
shifted material ($\dot{M}_{\rm ISM}>0$) -- in the face-on case it 
reaches about 50 M$_\odot$yr$^{-1}$ after the passage of the galaxy 
through the central cluster part. (Note that the amount of re-accretion 
is dependent on the extent of the evaluation zone.) 

\begin{figure}[t]
\centering
\includegraphics[height=0.48\textwidth,angle=270]{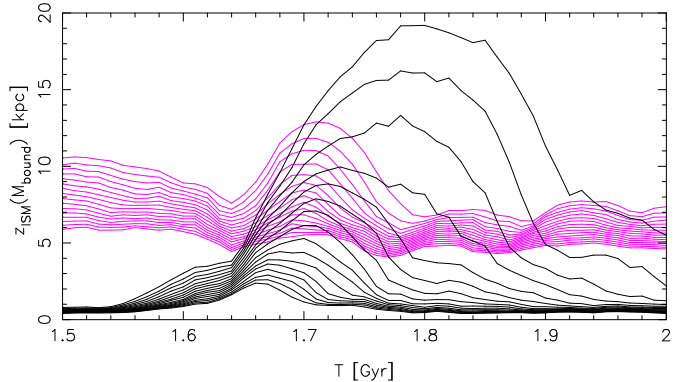}
\caption{\small 
ISM re-accretion: Distances behind the galaxy (in the direction of the 
wind) within which a certain fraction of actually bound ISM particles 
are enclosed (from 95~\% to 80~\%, {\it curves from top down}). Face-on 
({\it black}) and edge-on simulations of $R1\rho1$ run. (In the edge-on 
case the distances are taken in absolute value as initially one half of 
the particles occur ahead of the galactic mid-plane.)
}\label{zbound}
\end{figure}

The re-accretion process may be better followed in Fig.~\ref{zbound} 
which traces the actual distances of the bound ISM behind the galaxy (in 
the direction of the wind) for face-on and edge-on cases. In the 
face-on orientation, the ISM is by an increasing ram pressure first 
gradually pushed out of the disk plane, and then rapidly "thrown" to 
large distances, shortly after the cluster center passage. This material 
returns to the disk at later simulation times. 
In the edge-on orientation, the disk is first compressed by the rising 
ram pressure, and then some material is shifted to larger distances. 
The peak in Fig.~\ref{zbound} is, however, much smaller than in the 
face-on case. 
The reason why in near edge-on orientations the pushed material stays 
close to the disk is in the conservation of the angular momentum of the 
ISM particles rotating in the disk. When pushed by the ram pressure, 
the particles wind up around the disk. We checked this behavior in a 
more efficient run $R4\rho1$ where almost the same amount of ISM is 
stripped as in the standard face-on run. 

Apart from the total stripped amount, no radical differences in the 
duration of the stripping process of variously tilted galaxies can be 
seen from our simulations. Several details, like a slight time shift of 
the maximum of the stripping rate, or the course of re-accretion, look 
different in edge-on and face-on orientations, however these effects 
seem to be of secondary nature.

\subsection{Static vs. dynamic ICM pressure}
\begin{figure}[t]
\centering
\includegraphics[height=0.48\textwidth,angle=270]{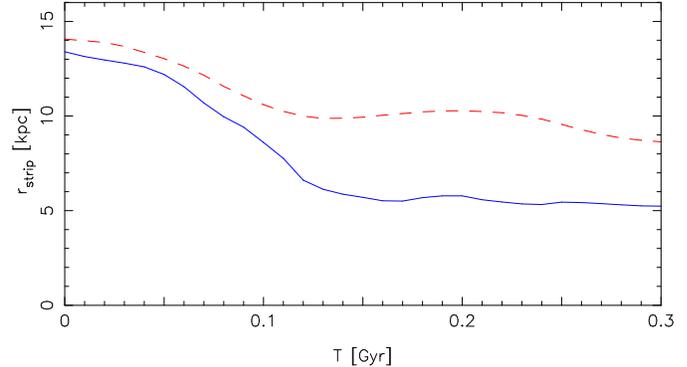}
\caption{\small 
Ram pressure stripping vs. zero-order test: Actual estimate of the 
stripping radius of the face-on galaxy in the standard $R1\rho1$ 
cluster ({\it solid}) as a function of time, vs. radius of the disk 
confined by static ICM pressure ({\it dashed}, the 95~\% radius from 
Fig.~\ref{zero_test}). 
}\label{calib}
\end{figure}

How does the effect of the disk radius reduction due to ram pressure 
compete with the disk confinement due to the static ICM pressure 
discussed in Section~\ref{zero}? In Fig.~\ref{calib} the rate of the 
disk radius reduction caused by the ram pressure stripping in the 
standard face-on run is compared with the upper curve of 
Fig.~\ref{zero_test}. It shows that the effect of the ram pressure is 
more important, even though the zero-order test took place in the 
center of the standard cluster where the disk experienced the maximum 
external pressure possible. In the ram pressure simulation the 
in-flying galaxy enters the ICM distribution at the truncation radius 
where the ICM density (and its pressure) is only small, and 
plunges deeper towards the cluster center, which is crossed very fast. 
Thus the actual role of the static ICM pressure will be even less than 
in Fig.~\ref{calib}.

Although the disk confinement due to the static ICM pressure operates 
against ram pressure -- it does not strip the gas but makes the disk 
more resistant to the ram pressure -- the tendency of the increasing 
effect of the static ICM pressure in "larger" clusters goes in the same 
direction as the effect of ram pressure stripping. Recently, 
\citet{tonnesenbryan09} have taken into account the multi-phase nature 
of the gas disk and found that the stripping radius is considerably 
smaller for galaxies with radiative cooling, although the amount of 
stripping is similar to that in galaxies without cooling. This goes in 
the same direction as the above effect observed in our simulations. 

We conclude that the disk confinement due to static external ICM 
pressure is much less important than the stripping due to ram pressure 
and we neglect this effect in analyzing the simulations in the 
following sections.

\begin{table*}[t]
\centering \caption{\small List of performed simulations -- results.
From left to right: fraction of the stripped ISM mass for simulations
with decreasing inclination, stripped mass fraction estimate of GG72
criterion, corresponding stripping radii and the estimate of GG72.}
\begin{tabular}{lccccccccccccc}
\hline \hline \rule{-3.pt}{2.6ex} run & $M_{\rm strip}^{90^\circ}$ &
$M_{\rm strip}^{70^\circ}$ & $M_{\rm strip}^{45^\circ}$ & $M_{\rm
strip}^{20^\circ}$ & $M_{\rm strip}^{0^\circ}$ & $M_{\rm strip}^{\rm
GG72}$ & $r_{\rm strip}^{90^\circ}$ & $r_{\rm strip}^{70^\circ}$ &
$r_{\rm strip}^{45^\circ}$ & $r_{\rm strip}^{20^\circ}$ & $r_{\rm
strip}^{0^\circ}$ & $r_{\rm strip}^{\rm GG72}$\\
& (\%) & (\%) & (\%) & (\%) & (\%) & (\%) & (kpc) & (kpc) &
(kpc) & (kpc) & (kpc) & (kpc)\\
\hline \rule{-3.pt}{2.6ex}
$R4\rho8$& 93 & 91 & 89 & 86 & 70 & 93 & 1.5 & 0.9 & 1.4 & 0.8 & 1.2 & 1.5\\
$R4\rho4$& 85 & 85 & 80 & 72 & 52 & 83 & 1.6 & 1.8 & 1.6 & 1.6 & 2.0 & 2.4\\
$R4\rho1$& 59 & 56 & 48 & 34 & 24 & 57 & 2.5 & 2.6 & 2.6 & 3.7 & 5.1 & 4.5\\
$R4\rho0$& 26 & 25 & 18 & 10 &  6 & 36 & 6.0 & 6.2 & 6.5 & 6.5 & 7.1 & 6.8\\
 \rule{-3.pt}{3.6ex}
$R1\rho8$& 79 & 77 & 70 & 54 & 36 & 92 & 1.5 & 1.6 & 1.6 & 2.2 & 2.7 & 1.6\\
$R1\rho4$& 63 & 61 & 50 & 30 & 19 & 81 & 2.1 & 2.5 & 3.1 & 3.2 & 4.0 & 2.5\\
$R1\rho1$& 30 & 27 & 20 & 10 &  5 & 57 & 5.5 & 6.3 & 6.5 & 7.2 & 8.0 & 4.5\\
$R1\rho0$&  9 &  8 &  6 &  2 &  1 & 36 & 8.6 & 9.0 & 9.3 & 9.9 & 10.3 & 6.8\\
 \rule{-3.pt}{3.6ex}
$R0\rho8$& 37 & 35 & 27 & 13 &  7 & 92 & 5.5 & 4.6 & 6.1 & 6.5 & 7.2 & 1.6\\
$R0\rho4$& 22 & 21 & 14 &  6 &  3 & 81 & 6.8 & 7.7 & 8.0 & 9.2 & 9.2 & 2.5\\
$R0\rho1$&  5 &  5 &  3 &  1 &  1 & 57 & 9.8 & 9.9 & 10.2 & 10.6 & 10.7 & 4.5\\
$R0\rho0$&  1 &  1 &  0 &  0 &  0 & 36 & 11.1 & 11.1 & 11.2 & 11.4 & 11.5 & 6.8\\
\end{tabular}
\label{tab_results}
\end{table*}

\section{Results}\label{sec:res}
In this section the results of the basic set of simulation runs 
introduced in Table ~\ref{tab_runs} are described. For each run listed 
we have performed five simulations with increasing disk inclination: $i 
= 0^\circ$ (edge-on), 20$^\circ$, 45$^\circ$, 70$^\circ$ and 90$^\circ$ 
(face-on). The overall Table~\ref{tab_results} provides the results: 
mass fraction of the stripped material ($M_{\rm strip}$) at the end of 
the simulation for five disk inclinations, stripped mass fraction 
estimated with the GG72 formula, stripping radii $r_{\rm strip}$ for five 
inclinations, and the stripping radius corresponding to GG72.

\subsection{Stripped amount}
\begin{figure}[t]
\centering
\includegraphics[height=0.48\textwidth,angle=270]{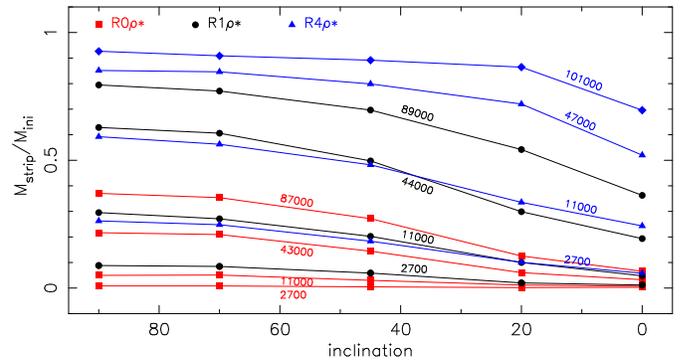}
\caption{ \small
Stripped mass fraction as a function of disk inclination for runs from
Table~\ref{tab_results}. Each run is labeled with the peak value of
the ram pressure. The inclination is 0$^\circ$ for edge-on and 
90$^\circ$ for face-on orientation.}
\label{mstrip_inclin}
\end{figure}

The fraction of the ISM particles that is no longer bound to the
galaxy at the final simulation time yields the stripped mass fractions,
whose dependence on inclination is shown in Fig.~\ref{mstrip_inclin}.
The general tendency of the plot is evident -- with the inclination
decreasing from face-on to edge-on, the amount of stripping declines.
Furthermore:
\begin{itemize}
\item Previous simulations of \citet{quilis00},
\citet{roedigerbruggen06} and others have shown that there is almost no
difference between $i=70^\circ$ and face-on stripping ($i=90^\circ$). This is true in
Fig.~\ref{mstrip_inclin} for all runs.
\item For large pressure peaks (runs $R4\rho4$ and $R4\rho8$), the
stripping amount is almost independent of the inclination (for angles
higher than about 20$^\circ$).
\item The dependence of the stripping amount on inclination is more
pronounced for smaller ram pressure peaks.
\item Runs with the same value of $R_{\rm c,ICM} \cdot \rho_{\rm
0,ICM}$ quantity show close profiles of the $M_{\rm strip}(i)$ curves
-- e.g. ($R4\rho1$ and $R1\rho4$), or ($R4\rho0$, $R1\rho1$, and
$R0\rho4$).
\end{itemize}

\begin{figure}[t]
\centering
\includegraphics[height=0.48\textwidth,angle=270]{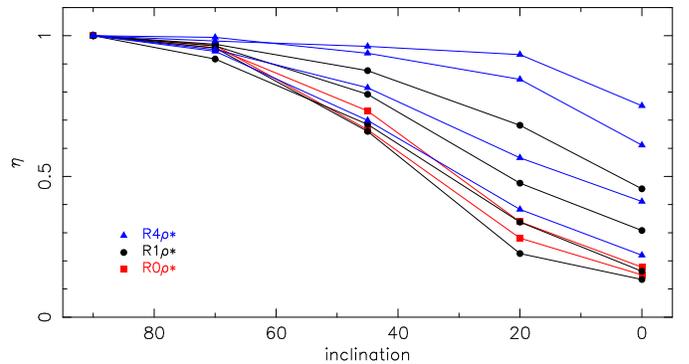}
\caption{ \small
Stripping efficiency with respect to inclination $\eta=M_{\rm 
strip,i}/M_{\rm strip,90^\circ}$ as a function of the inclination (runs 
$R0\rho1$ and $R0\rho0$ are not plotted).}
\label{eff_inclin}
\end{figure}

The behavior with varying inclination is more clearly displayed in
Fig.~\ref{eff_inclin} which shows a new parameter $\eta$ that we
name the {\it stripping efficiency with respect to inclination}. It is 
the ratio of the stripped mass fraction relative to the face-on case:
\begin{equation}\label{eq_eta}
\eta (i) = \frac{M_{\rm strip,\it i}}{M_{\rm strip, 90^\circ}} =
\frac{1-M_{\rm fin,\it i}/M_{\rm ini}} {1-M_{\rm
fin,90^\circ}/M_{\rm ini}}.
\end{equation}
Thus, $\eta$ characterizes the relative strength of a given ram
pressure profile to strip the ISM from an inclined galaxy with
respect to the face-on case. In Fig.~\ref{eff_inclin}, one observes 
that the stripping efficiency always declines with decreasing 
inclination angle. Moreover, both wider and higher ram pressure peaks 
yield higher efficiencies. For smaller pressure peaks, $\eta$ falls
fastest at medium inclinations (between 20$^\circ$ and 70$^\circ$),
while for large peaks at inclinations smaller than $20^\circ$.

\begin{figure}[t]
\centering
\includegraphics[height=0.48\textwidth,angle=270]{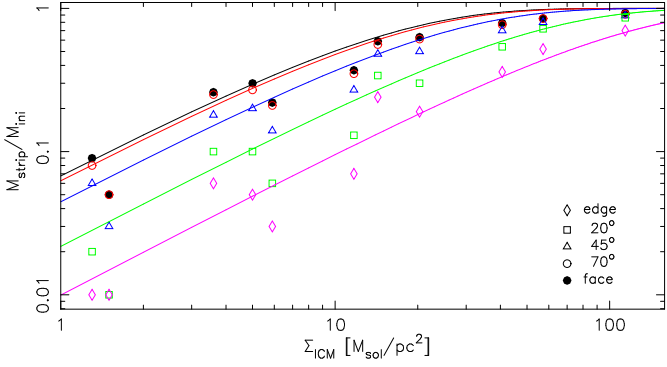}
\caption{ \small
Stripped mass fraction as a function of the column density of the
encountered ICM. The symbols denote the results of the SPH simulations
for various inclinations. The curves correspond to the fitting curve in
Eq.~\ref{fit}.
} \label{Mstrip_SigICM}
\end{figure}

\begin{figure}[t]
\centering
\includegraphics[height=0.48\textwidth,angle=270]{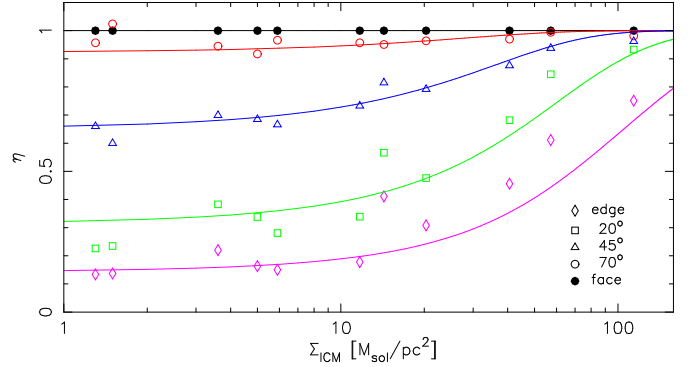}
\caption{ \small
Stripping efficiency with respect to inclination as a function of 
the column density of the encountered ICM. The curves correspond to the 
fitting curve in Eq.~\ref{fit}.
} \label{eff_SigICM}
\end{figure}

Stripping of face-on galaxies depends on the amount of the ICM 
encountered along the trajectory \citep[Fig.~15 in][]{jachym07}: 
galaxies crossing clusters with such combinations of $R_{\rm c,ICM}$ 
and $\rho_{\rm 0,ICM}$ parameters that $\Sigma_{\rm ICM}\sim \rm const$ 
were stripped to the same level. For inclined cases this hypothesis 
is tested in Figs.~\ref{Mstrip_SigICM} and \ref{eff_SigICM}, which show 
the stripped mass fraction and the stripping efficiency with respect to 
inclination as a 
function of the column density of the encountered ICM, respectively. 
Different colors refer to simulations with different inclination 
angles. The following trends are apparent
\begin{itemize}
\item with increasing amount of encountered ICM the stripped mass
fraction and the efficiency increase;
\item there is a tendency for the results from each inclination to
follow a certain close relationship, despite some systematic deviations
due to the different ICM density distributions;
\item for high $\Sigma_{\rm ICM}$, these relations saturate towards
complete stripping;
\item for lower ICM column densities, edge-on stripping is reduced
with respect to face-on by a constant factor.
\end{itemize}

All these results indicate that the column density of the ICM
encountered during the ISM-ICM interaction is the key parameter
determining the stripping outcome. It is much more important than the
maximum value of the ram pressure experienced along the orbit (GG72 
criterion).

The numerical results in Fig.~\ref{Mstrip_SigICM} can be fitted with
the following formula
\begin{equation}\label{fit}
\frac{M_{\rm strip}}{M_{\rm ini}} = 1 - \exp\left[-0.01\, \Sigma_{\rm
ICM} (1 + 6 \sin^{1.5} i)\right].
\end{equation}
The exponent 1.5 provides a rather satisfactory match to the results
with intermediate inclinations, as can be seen in both
Figs.~\ref{Mstrip_SigICM} and \ref{eff_SigICM}.

In Table~\ref{tab_results}, the stripped mass fraction estimated with 
the GG72 criterion is added. For the largest ram pressure peak ($R4\rho8$), 
the values match well the face-on results of the SPH simulations. For 
smaller peaks, however, GG72 overestimates the real stripped fraction. 
Of course, no dependence on the tilt angle is included in the GG72 
criterion, and its multiplication by a simple $\cos i$ term does not 
resolve the problem, as it yields zero stripping in edge-on cases.

\subsection{Stripping radii}
Due to an asymmetric shape of the gas disk that develops, especially
after an inclined stripping event, the determination of its final 
radius, i.e. the stripping radius, is not trivial 
\citep[see][]{roedigerbruggen06}. In close to edge-on cases,
all the stripped and shifted material stays in the plane of the disk
and makes the determination difficult.

In Table~\ref{tab_results}, the given values of stripping radii 
correspond to radii within which 90\% of the bound ISM is enclosed in 
the final simulation time. This may slightly underestimate the radius 
for the face-on orientation, while overestimating it for the edge-on case.
The values do not show a strong dependence for large and small peak
ram pressures; however, for medium pressure the trend of increasing
stripping radii towards edge-on orientations is visible.

\begin{figure}[t]
\centering
\includegraphics[height=0.48\textwidth,angle=270]{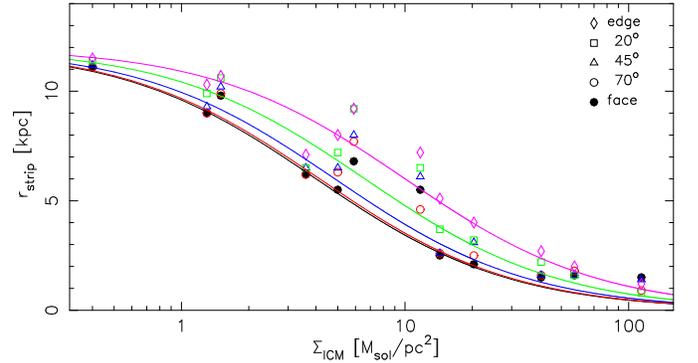}
\caption{ \small
Stripping radii $r_{\rm strip}$ versus column density of the 
encountered ICM. The symbols
are the same as in Fig.~\ref{Mstrip_SigICM}. The points are fitted with
the fitting function in Eq.~\ref{fit_rstrip}.
} \label{rstrip_inclin}
\end{figure}

As shown in Fig.~\ref{rstrip_inclin}, the stripping radii for all
simulations can be quite reasonably fitted with
\begin{equation}\label{fit_rstrip}
r_{\rm strip} = 12\, \mathrm{kpc}\, /\, [1 + \Sigma_{\rm ICM} (0.1 + 0.15
\sin i)].
\end{equation}

The stripping radii estimated in the last column of 
Table~\ref{tab_results} using the GG72 criterion show quite a good 
agreement for the largest run. However, there is a tendency to 
overestimate the SPH results. For smaller peaks, GG72 predicts less 
stripping corresponding only to the peak ram pressure.

\section{A simple test particle model}\label{sec:kin}
In order to study the influence of the parameters more efficiently, and 
to separate various effects -- the direct kinematics on the ISM induced 
by the ram-pressure, and the related hydrodynamical consequences -- we 
construct a very simplified model for the removal of gas from a
galaxy. The gas disk is represented by 10~000 point masses
distributed as in an exponential disk and held in the potential well
composed of the bulge, stellar disk, and dark halo, with parameters
identical to those used in the SPH computations. Initially, all
these particles are in circular orbits about the center of the
galaxy. Then, they are subjected to a Gaussian-shaped force pulse of
specified tilt angle with respect to the galactic plane, duration,
and maximum amplitude. The simulation starts at a time that is three 
times the pulse duration before the maximum. In the computation of the acceleration,
the inertial mass of each particle is taken to be proportional to
the gas surface density of the exponential disk, in order to account
for the correct computation of the ram pressure in that part of the
disk. For each particle, the equation of motion is integrated, and
when the total energy becomes positive and the particle becomes
unbound, it is marked and counted as having been stripped. If all
particles were to experience this force pulse, the fraction of
stripped particles is independent of the tilt angle, because all
reach escape velocity within time-spans that depend little on their
initial position in the galaxy and the force tilt angle.

In order to simulate the thickness of the disk, we introduce a
shielding or shadowing of the particles downstream of those
particles facing the force. In this, we follow the modeling by
sticky particles by \citet{vollmer01a}. Particles that are
shielded by at least one upstream particle are assumed to experience
only the galaxy's internal forces. Experiments show that with the
choice of a suitable value for the shadowing radius (typically 200~pc), 
we are able to reproduce well the amount of stripping for the
standard model of the SPH computations, as well as its dependence on
the tilt angle.

\begin{figure}[t]
\centering
\includegraphics[clip=true,bb=73 72 543 745,height=0.48\textwidth,angle=270]{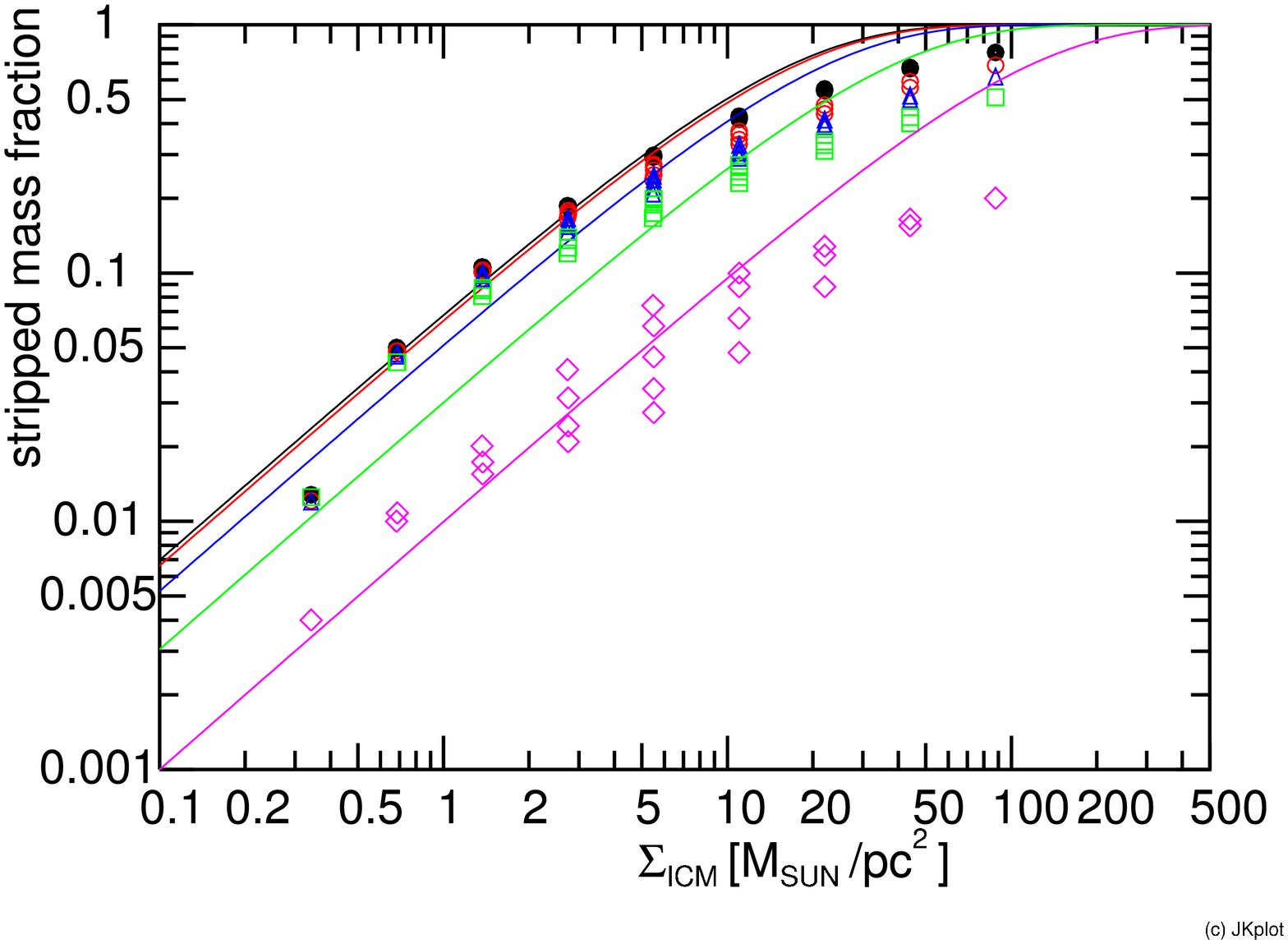}
\caption{\small 
The dependence of the stripped mass fraction on the column density of 
the swept-up ICM, for different tilt angles. The dots are results from 
a grid of simplified point-mass models, covering a range from 1/16 to 
16 times in both duration and strength of the ram pressure force pulse, 
centered about the value for our standard model. The curves are the 
analytical fit to the SPH results. The symbols and colors are identical 
to those of Fig.~\ref{Mstrip_SigICM}. Note that the range of parameters 
in both axes is much wider.
}\label{FigA}
\end{figure}

As is seen in Fig.~\ref{FigA}, the stripped mass fractions from these 
simplified models behave in a very similar way as the SPH results. 
While at high values of the ICM column density the stripping is less 
severe and the relation is slightly flatter, the influence on the tilt 
angle is nearly identical -- one obtains the same amount of stripping in 
the edge-on configuration as in the face-on case for an ICM column density 
that is larger by about a factor of 6. This indicates that this modeling 
approach, and hence also the sticky-particle computations by 
\citet{vollmer01a}, appears to be quite reliable in capturing the main 
features of ram-pressure stripping. However, this simplified approach 
has notable differences to the full hydrodynamical approach, in 
features that we consider below.

\begin{figure}[t]
\centering
\includegraphics[clip=true,bb=73 72 543 745,height=0.48\textwidth,angle=270]{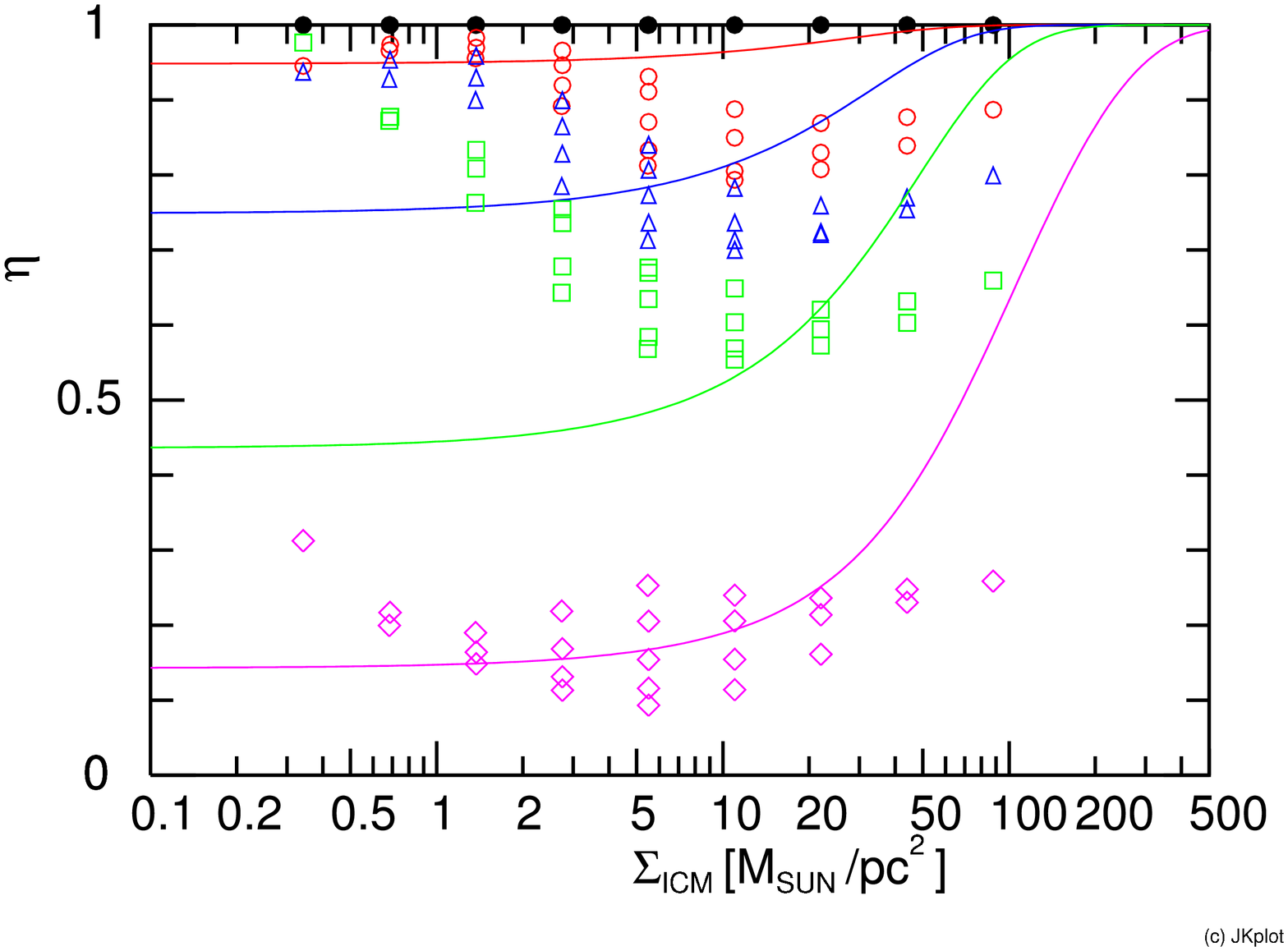}
\caption{\small
Similar to Fig.~\ref{FigA}, but for the dependence of the stripping 
efficiencies with respect to inclination as a function of the 
column density of the encountered ICM.
}\label{FigB}
\end{figure}

The stripping efficiencies with respect to inclination (Fig.\ref{FigB}) 
show that the ratios of face-on and edge-on results are as large as 
from the SPH models, but that there are some limitations. First 
of all, at intermediate angles the simplified modeling gives results 
that are closer to the face-on case than we find in SPH. Furthermore, 
at high ICM column densities, one does not find such a steep increase of
stripping as experienced in SPH. We searched quite extensively in
parameter space, but could not find situations where the behavior
matched the SPH results more closely. Evidently, this is a
consequence of the limitations of the simplified treatment. We take
this as an indication that this kind of diagram is sensitive to the
different treatment of the (hydro-)dynamics of the problem.

The simplified modeling does allow us to separate the different 
physical phenomena in the ram-pressure process. In particular it helps 
to identify the hydrodynamical shielding by the matter lying
upstream as the principal cause for the dependence of stripping on
tilt angle, seen in our SPH results. The deviations seen with
respect to the SPH results suggest that the details of the behavior 
changes with tilt angle comes from the finer details of the 
hydrodynamics involved in the problem.

\section{Comparison with the analytical criterion}\label{sec:analyt}
\begin{figure}[t]
\centering
\includegraphics[height=0.48\textwidth,angle=270]{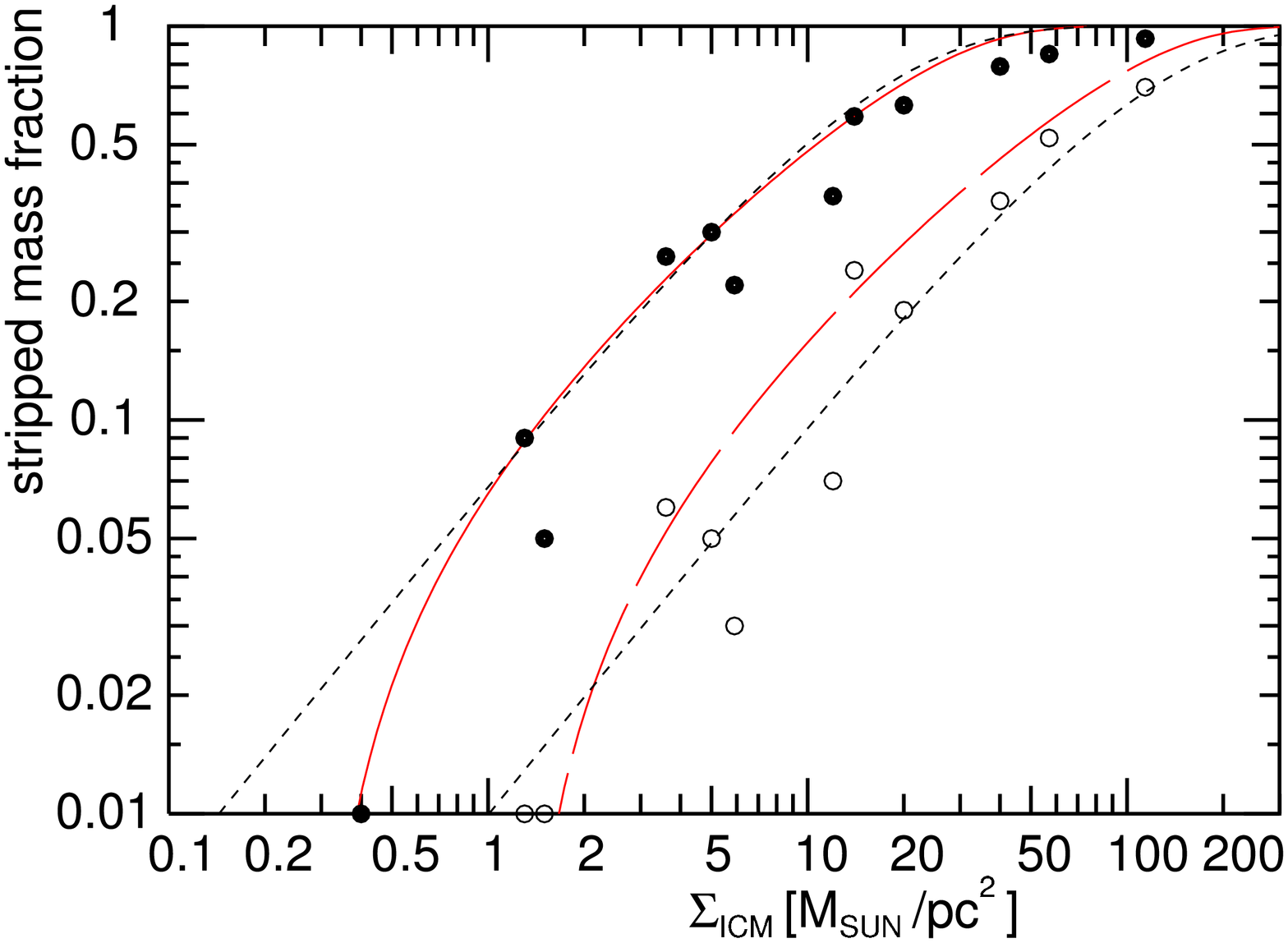}
\caption{\small
The relation between stripped mass fraction and the column density of
the swept-up ICM gas, computed for our standard model from our
analytical criterion. The dots are results from the SPH models for the 
face-on configuration, compared to the predictions, adopting a value of 
$A=0.5$ ({\it solid curve}). The open circles are results in edge-on 
orientation, and the dashed curves are the predictions with an adopted 
value $A=0.12$. The dotted curves come from the fitting formula 
Eq.~\ref{fit}.
}\label{FigC}
\end{figure}

From the consideration of the momentum provided by the in-streaming ICM 
to the gas in the galactic disk \citet{jachym07} derived the
condition under which a volume element of disk gas (with column density
$\Sigma_{\rm ISM}$ for the face-on situation) would escape from a 
galaxy,
\begin{equation}
\langle v\rangle_{\rho_{\rm ICM}} \Sigma_{\rm ICM} / \Sigma_{\rm ISM} = 
v_{\rm after} > v_{\rm esc} = \sqrt{-2\Phi(r)},
\end{equation}
i.e. if the speed $v_{\rm after}$ exceeds the local escape speed. The
stripping radius would then be defined as the radius where $v_{\rm
after}$ = $v_{\rm esc}$ and the stripped mass fraction as the mass
exterior to that radius:
\begin{equation}
M_{\rm strip} = \int_{r_{\rm strip}}^{r_{\rm out}} 2 \pi \Sigma_{\rm
ISM}(r)\, r\, dr \ / \ M(r_{\rm out}),
\end{equation}
where $r_{\rm out}$ is the truncation radius of the disk.
For an exponential disk $\Sigma_{\rm ISM}(r) = \Sigma_0 \exp(-r/a_d)$
with $M_{\rm ISM} = 2\pi \Sigma_0 a_d^2$ we have
\begin{equation}
M_{\rm strip} = \frac{(a_d+r_{\rm out})\, e^{-r_{\rm out}/a_d} -
(a_d+r_{\rm strip})\, e^{-r_{\rm strip}/a_d}} {(a_d+r_{\rm out})\,
e^{-r_{\rm out}/a_d}-a_d}.
\end{equation}
On the other hand, the escape velocity as a function of galactocentric 
radius is the sum of the contributions from all components, i.e. bulge, 
disk, and dark halo. At the stripping radius the escape velocity equals 
the speed $v_{\rm after}$, which is a function of the ICM column 
density which we shall now take as the independent variable. Hence we 
can compute the stripping radius as a function of ICM column density, 
and with that we get the stripped mass fraction. This is easily 
evaluated numerically by computing for each galactocentric radius the 
mass fraction outside of that radius, and on the other hand the 
quantity
\begin{equation}
\Sigma_{\rm ICM} = v_{\rm esc} \frac{\Sigma_{\rm ISM}(r)} {\langle 
v\rangle_{\rho_{\rm ICM}}} = A \sqrt{-2 \Phi(r)}\, \frac{\Sigma_{\rm 
ISM}(r)} {\langle v\rangle_{\rho_{\rm ICM}}}.
\end{equation}
Figure \ref{FigC} depicts the relation between $M_{\rm strip}$ and
$\Sigma_{\rm ICM}$ for our standard model. The above equation is able
to match the SPH results very closely, if we adopt $A=0.5$. In this 
diagram, our fit formula to the SPH results (Eq.~\ref{fit}) gives a 
function very similar in character to the one derived 
from our analytical criterion. The choice of the outer disk radius 
influences the shape of the curve in Fig.~\ref{FigC}; we find that a 
value of 20 kpc gives a better match than the value of 16 kpc assumed 
for the SPH models. This detail is not incorporated in the fit formula, 
whose results therefore match the SPH results at low stripping less 
well.

Although the SPH results from the edge-on case do exhibit a larger
scatter than the face-on data, it is quite remarkable that they can 
still be matched by the analytical prediction for the face-on case, if 
one simply increases the necessary ICM column density. This means that 
the edge-on models can well be considered as having a lower stripping
efficiency, with a reduction factor of about 5. This factor may be
regarded as depending only on the tilt angle $i$, as
\begin{equation}
A(i) = 0.12 + 0.38 \sin i.
\end{equation}
A sine-type dependence could be understandable as a geometrical 
projection effect, in that the edge-on galaxy presents a smaller cross 
section to the ICM. If it were such a pure effect of the effective cross 
sections, one should expect for our standard model a ratio of the areas 
of $4/0.25 = 16$, from the disk radial scale being 4 kpc and its height 
being 0.25 kpc. For the moment, we prefer to take our analytical formula 
as a convenient expression which encapsulates essential features of the
problem rather than as giving a complete description.

\section{Discussion}\label{sec:disc}
We compare our results to the findings of previous numerical studies 
which treated inclined ram pressure stripping. These investigations 
differ in their numerical treatment of hydrodynamics, the model for the 
galaxy, and the way the ICM gas flows around the galaxy. 

\subsection{General trends}
\begin{figure}[t]
\centering
\includegraphics[height=0.48\textwidth,angle=270]{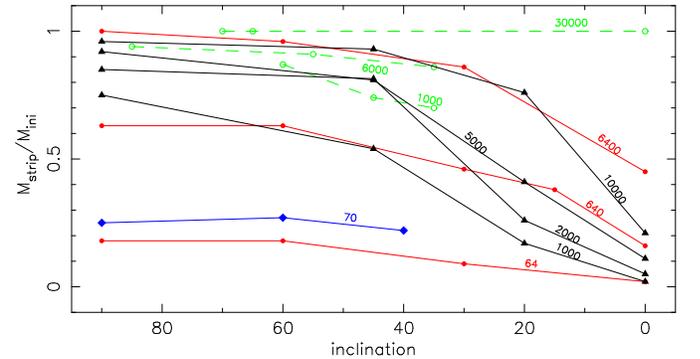}
\caption{\small Stripped mass fraction as a function of the disk
inclination, from \citet{vollmer01a} - {\it black}, 
\citet{roedigerbruggen06} - {\it red}, \citet{roedigerbruggen07} -
{\it green}, and \citet{schulzstruck01} - {\it blue}. Labels give 
values of the peak ram pressure, in units of cm$^{-3}$(km/s)$^2$. For 
\citet{roedigerbruggen07} the actual values of the inclination are 
given for the instant of pericenter passage.
}\label{compare}
\end{figure}

The fraction of the ISM mass that is stripped in the simulations of
\citet{vollmer01a}, \citet{roedigerbruggen06},
\citet{roedigerbruggen07}, and \citet{schulzstruck01} is plotted in
Fig.~\ref{compare} as a function of inclination angle. 
The curves join runs with different disk inclinations, and 
are labeled by the maximum ram pressure. It must be stressed that for 
different curves not only does the maximum value of the ram pressure 
vary, but also its duration. These general features are apparent in 
Fig.~\ref{compare}:
\begin{itemize}
\item for inclinations larger than about 50$^\circ$ there is very 
little dependence of the stripped fraction;
\item towards the edge-on orientation, on the contrary, the amount of
stripping drops strongly; 
\item the curves flatten for higher peak ram pressures.
\end{itemize}

Our results show the same features (cf. Fig.~\ref{mstrip_inclin}). A 
closer comparison shows that the near edge-on models by 
\citet{vollmer01a} are substantially more strongly stripped than ours 
for the same peak ram pressure. However, since their Fig.~7 suggests 
that the radial scale-length of the gas disk is longer than in our 
model, one should expect that due to the higher ISM surface density 
they should be stripped less severely. This suggests that the 
shielding mechanism used by \citet{vollmer01a} is overly effective in 
near edge-on orientations and thus overestimates the stripping, at 
least in comparison to the full hydrodynamic treatment of 
\citet{roedigerbruggen06} and our SPH approach. 

For the same values of the stripped mass fraction, the simulations by 
\citet{roedigerbruggen06} require a substantially lower value of the 
peak ram pressure than our models or those by \citet{vollmer01a}, 
because \citet{roedigerbruggen06} consider the flow of the ICM 
to be constant in time, hence much longer than in the other models. 

\begin{figure}[t]
\centering
\includegraphics[height=0.48\textwidth,angle=270]{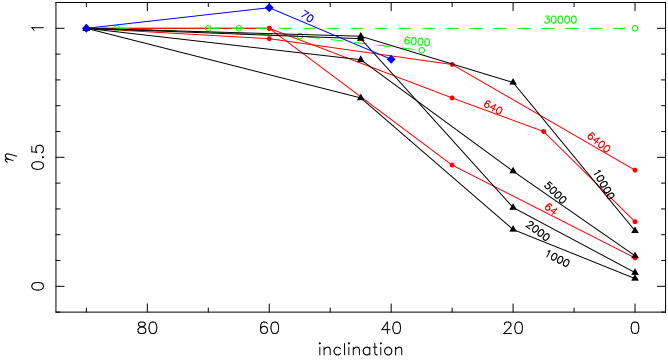}
\caption{\small Stripping efficiency with respect to inclination as a 
function of the angle for the same results as in Fig.~\ref{compare}.
}\label{compare_eta}
\end{figure}

The stripping efficiency with respect to inclination ($\eta$) measures 
the relative stripping 
potential of individual inclinations with respect to the face-on case. 
It might thus serve as a good instrument for comparing results from 
different numerical codes. From Fig.~\ref{compare_eta} it is obvious 
that the results of different authors give very similar dependences 
on the inclination angle. In spite of the basic difference in the 
absolute values for the maximum ram pressure, the results of both 
\citet{roedigerbruggen06} and \citet{vollmer01a} show that higher ram 
pressures result in a lower dependence of the stripping on inclination 
angle, i.e. making the inclined stripping as effective as the face-on 
case. Thus, the features of the inclination dependence, which our 
present study shows in a systematic way, has already been seen 
in earlier studies.

\subsection{Vollmer's fitting formula}
\begin{figure}[t]
\centering
\includegraphics[height=0.48\textwidth,angle=270]{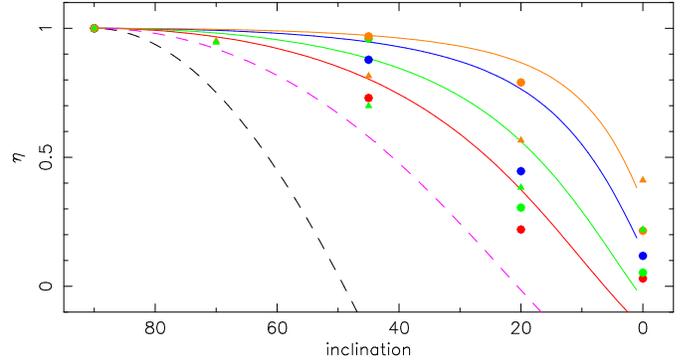}
\caption{ \small Curves of stripping efficiency with respect to 
inclination $\eta$ as a
function of tilt angle as calculated from the fitting formula
of \citet{vollmer01a} for various values of the peak ram
pressure: 100, 300, 1000, 2000, 5000 and 10~000 cm$^{-3}$(km/s)$^2$
({\it from left to right}). Results of Vollmer's simulations
({\it dots}) correspond to the solid curves. Results of our
$R4\rho0$ and $R4\rho1$ runs with pressure peaks of $\sim 2700$ and 
11~000 cm$^{-3}$(km/s)$^2$ are added ({\it triangles}).
} \label{vollmer_fit}
\end{figure}

\citet{vollmer01a} derived from their simulation results a formula
which relates the fraction of the final and initial total gas mass
to the peak ram pressure and inclination angle (their Eq.~21):
$M_{\rm ini}/ M_{\rm fin} = 0.25\, p_{ram}^{max}\, \sin^2 [0.9\,
(i+10^\circ)] + 0.84$, where $p_{ram}^{max}$ is normalized by
$p_0=100$ cm$^{-3}$ (km/s)$^2$. Using this fitting formula we
reproduce Vollmer's Fig.~17 in Fig.~\ref{vollmer_fit}: as a
function of angle we plot the stripping efficiency with respect to 
inclination $\eta$. It shows that although the fitting function does 
not match the simulations perfectly (and even gives for low ram 
pressures negative values of $\eta$) the general trend corresponds well 
to the numerical results, and to our findings as well: that for high 
ram pressures the efficiency of stripping is high and less dependent on 
the tilt angle than for low ram pressure values.

\subsection{Dependence on ICM column density}
\begin{figure}[t]
\centering
\includegraphics[height=0.48\textwidth,angle=270]{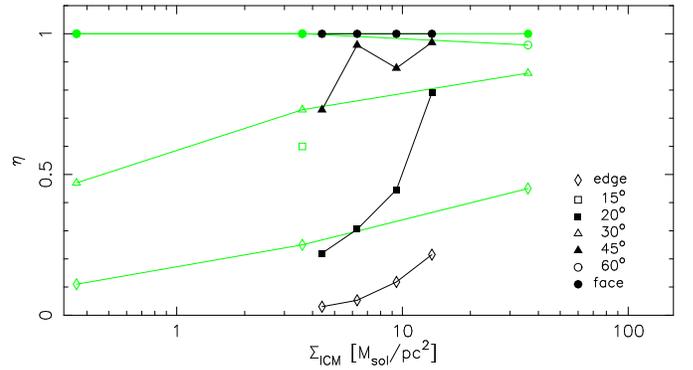}
\caption{\small 
Stripping efficiency with respect to inclination $\eta$ as a function 
of the column density of the encountered ICM for simulations of 
\citet{roedigerbruggen06} ({\it green}) and \citet{vollmer01a} ({\it 
black}).
}\label{compare_SigICM}
\end{figure}

To look for the dependence of the stripping results on $\Sigma_{\rm
ICM}$ in the previous simulations, we estimate these values from the 
parameters of their runs. In \citet{roedigerbruggen06}, a constant flow
of ICM with three values of density (0.01, 0.1, and 1 $\times$
10$^{-26}$g$\,$cm$^{-3}$) provokes a wind at 800 kms$^{-1}$ lasting for 
1 Myr. This yields values of the ICM column density of
0.4, 4, and 36 $M_\odot$pc$^{-2}$ for their weak, medium, and strong
pressure models, respectively. \citet{vollmer01a} approximate ram 
pressure profiles with Lorentz functions, whose duration is determined 
by the condition that 500~Myr before the peak, a constant pressure of 
50 cm$^{-3}$(km/s)$^2$ is reached. For their four peak ram pressures of 
1000, 2000, 5000, and 10~000 cm$^{-3}$(km/s)$^2$ we derive widths 
$t_{\rm HW}=$ 115, 80, 50, and 35 kpc, respectively, which yield values 
of $\Sigma_{\rm ICM}$: 4.4, 6.2, 9.4, and 13.5 $M_\odot$pc$^{-2}$.

Figure~\ref{compare_SigICM} compares the $\eta(\Sigma_{\rm ICM})$
dependence for the results of \citet{vollmer01a} and 
\citet{roedigerbruggen06}. We do not include the results of 
\citet{roedigerbruggen07}, as the tilt angle changes along the galaxy's 
orbit. The results of both \citet{vollmer01a} and 
\citet{roedigerbruggen06} show the same trend as that revealed in our 
simulations: with increasing $\Sigma_{\rm ICM}$ the inclination 
efficiency grows. It is worth noting that the increase is much steeper 
in the case of \citet{vollmer01a}, in comparison to 
\citet{roedigerbruggen06} and our own results. Since the parameters of 
the model galaxies are rather similar and the stripping efficiency 
$\eta$ measures the relative potential of inclined stripping with 
respect to the face-on case, it could well be that these differences 
are due to the different ways the hydrodynamics is treated.

\subsection{Limitations of strictly radial orbits}
To separate the effects of the disk tilt angle on the stripping results 
we have focused in this paper on galaxies on strictly radial orbits. 
Then the tilt angle stays constant along the orbit. In more realistic 
orbits however, the disk inclination usually changes with time. Thus, 
almost all galaxies are likely to experience once along their orbit 
an efficient (close-to) face-on ram pressure phase. Its duration and 
strength is then determined by the parameters of the orbit. 
As the results of our simulations have shown, the tilt-variance might 
be important for situations with compact ICM distributions where the 
face-on stripping really is more efficient than the tilted ones. In 
extended clusters, however, the dependence on the disk inclination 
vanishes as the stripping tends to saturate.

Concerning the stripping potential of individual models, note that the 
strictly radial orbits used in our calculations may be considered as 
being equivalent to slightly elliptical ones: The ram pressure profile 
along a strictly radial orbit through a given ICM distribution may well 
be the same as the ram pressure profile along an elliptical orbit, however 
through a modified ICM distribution that is more centrally peaked 
(higher) and less extended (more compact). In other words, on an 
elliptical orbit the galaxy passes through the slope of the ICM density 
peak.

\section{Conclusion}
We have studied in detail the influence of disk tilt angle on the ram 
pressure stripping efficiency in galaxy clusters. A grid of N-body/SPH 
numerical simulations, varying the density of the ICM, the peak 
ram-pressure, its duration, and the orientation of the galaxy was 
performed. We summarize below the main results of the simulations:
\begin{itemize}
\item the stripping amount is lower for galaxies moving edge-on 
with respect to the ICM wind, in agreement with previous 
simulations;
\item apart from the total stripped amount, there is no 
radical difference in the duration of the stripping process of 
variously tilted galaxies;
\item due to conservation of angular momentum, ram pressure shifts 
the ISM in near-edge-on orientations to smaller distances behind the 
galaxy than in the face-on case;
\item the dependence of the stripping amount on the disk tilt angle 
vanishes for large clusters, where the amount of the encountered 
ICM ($\Sigma_{\rm ICM}$) is large;
\item on the other hand, for galaxies crossing small ICM overdensities 
or debris structures left over in the cluster from recent stripping 
events, the dependence on the tilt angle is more pronounced;
\item we emphasize the role of the amount of the encountered ICM 
to the ram pressure stripping results;
\item the stripping efficiency with respect to inclination $\eta$ (i.e. 
the ratio of the stripped mass fraction relative to the face-on case) 
depends strongly on the disk tilt angle for small ram pressure peaks 
but again vanishes for large values of $\Sigma_{\rm ICM}$ when 
stripping saturates;
\item for non-face-on orientations, the sense of  rotation plays a 
role: contrary to expectations, it is not the side with highest ram 
pressure which is more stripped;
\item we propose a fitting formula (Eq.~\ref{fit}), which should now be 
tested against observations, deriving the ICM density from X-ray 
imaging of clusters, and the gas stripping from HI observations.
\end{itemize}

By comparison with a simple analytical stripping condition 
\citep{jachym07}, the total momentum imparted by the encountered ICM on 
the galaxy gas appears to be the main determinant of the stripping 
result. Contrary to the GG72 criterion or the fitting formula of 
\citet{vollmer01a} who assign the key role to the peak value of the ram 
pressure, we stress the importance of the dependence on the amount of 
the encountered ICM, together with the disk 
inclination angle: 
\begin{displaymath}
M_{\rm strip}/M_{\rm 
init} \sim \Sigma_{\rm ICM} (a+b\,\sin^\alpha i).
\end{displaymath}
The values of $a$, $b$, and $\alpha$ depend on the ISM distribution in 
the unperturbed disk. We use such a dependence to fit the outcome of 
our numerical simulations -- the stripped mass fraction and the 
stripping radius. The analytical stripping condition based on the 
momentum balance yields similar formulae. 

Using simple test particle models taking schematically into account the 
"shadowing" of the gas disk by the material occurring upstream, we can 
roughly obtain the main features of the stripping. The imparted momentum 
$\langle v\rangle_{\rho_{\rm ICM}} \Sigma_{\rm ICM}$ proves to be the 
key parameter of the ICM-ISM interaction. Such particle models however 
fail to reflect the true dependence of the stripping efficiency on ICM 
density and inclination revealed by the full hydrodynamics.

\begin{acknowledgements}
The authors gratefully acknowledge support by the Institutional 
Research Plan AV0Z10030501 of the Academy of Sciences of the Czech 
Republic, by the project 205/08/P556 of the Grant Agency of the Czech 
Republic, and by the Center for Theoretical Astrophysics (LC06014). We 
thank Volker Springel for valuable advice. The simulations have been 
carried out on the Ond\v rejov Cluster for Astrophysical Simulations 
(OCAS, Czech Republic) and the VIRGO computational cluster in Prague. 
We thank Jim Dale for English revision of the text. We would like to 
thank the referee for helping us to improve this paper considerably.
\end{acknowledgements}

\bibliographystyle{aa}
\bibliography{1469}
\end{document}